\documentclass[a4paper, 10pt]{article}
\usepackage{verbatim}
\usepackage{amsfonts}
\usepackage{amsmath}
\usepackage{a4wide}
\usepackage[english]{babel}
\usepackage[margin=1in]{geometry}
\usepackage{enumerate}
\usepackage{graphicx}
\usepackage{colortbl}
\usepackage{fixltx2e}
\usepackage{url}
\usepackage{caption}
\usepackage{subcaption}
\usepackage{wrapfig}
\definecolor{orange}{rgb}{1,0.5,0} 
\usepackage{bbding}
\usepackage{enumitem}
\setitemize{noitemsep}

\newcommand*\tick{\item[\textcolor{green}{\Checkmark}]}
\newcommand*\fail{\item[\textcolor{red}{\XSolidBrush}]}
\newcommand*\otick{\item[\textcolor{orange}{\Checkmark}]}
\newcommand*\ofail{\item[\textcolor{orange}{\XSolidBrush}]}

\title{Evaluating e-voting: theory and practice}
\author{Manon de Vries and Wouter Bokslag\\Department of Information Security Technology\\Technical University of Eindhoven}
\begin{document}
\maketitle
\thispagestyle{empty}

\begin{abstract}
In the Netherlands as well as many other countries, the use of electronic voting solutions is a recurrent topic of discussion. While electronic voting certainly has advantages over paper voting, there are also important risks involved. This paper presents an analysis of benefits and risks of electronic voting, and shows the relevance of these issues by means of three case studies of real-world implementations. Additionally, techniques that may be employed to improve upon many of the current systems are presented. We conclude that the advantages of E-voting do not outweigh the disadvantages, as the resulting reduced verifiability and transparency seem hard to overcome. %checked 
\end{abstract}
\newpage
\section{Introduction}
In the last decade, we have observed a shift from paper forms to their contemporary electronic equivalents. Many services that once required a citizen to physically present himself at a counter and fill out paper forms have now been made available over the internet, where, after some kind of authentication, the same service is provided over a digital medium. The advantages are numerous and significant, as it is more convenient for the citizen and digital information is far more suitable for automated processing. \\
\\
As a consequence, it is only fair that one starts thinking about an electronic equivalent for voting. Most countries still use paper ballots which are counted by hand after the voting period ends. This obviously has drawbacks: paper is wasted, manual vote counting takes time and is potentially more error-prone than electronic vote counting. As tempting as electronic voting may seem, it is important to realize the potential risks and drawbacks. The possibility to cast a vote, and the confidence that all votes are being taken into account in a honest manner is one of the main pillars of a modern democracy. The trade-off between advantages and risks of electronic voting is thus one that deserves careful thought. \\
\\
In section \ref{sec:riskbenefit}, the different possible advantages and disadvantages electronic voting solutions may have will be discussed. Due to different design choices, not all implementations share the same advantages and risks. Different real-world implementations of electronic voting systems and their properties will be discussed in section \ref{sec:casestudies}, where three case studies are presented. As there are approaches to electronic voting that have not yet been implemented in large-scale real-world solutions, we will discuss potential future improvements over current implementations in section \ref{sec:improvements}. We will conclude the paper with section \ref{sec:conclusion}, where, based on the information and reasoning presented in the paper, we draw conclusions about the acceptability of current and, to some extent, future electronic voting solutions. 

\subsection{Definition}
When considering electronic voting solutions, a clear distinction can be made between what is commonly referred to as \emph{I-voting} or \emph{internet voting} and a more traditional form of electronic voting that employs controlled voting booths where the actual vote is cast on an electronic device. In this paper we will discuss both regular e-voting and I-voting. \\
\\%checked
For the sake of completeness, it must be noted that there is also a category of electronic voting solutions that remain very close to the concept of a physical, paper ballot. These systems are mainly an electronic interface to generate paper votes, which may then be counted more efficiently. Different approaches exist, such as punch cards, electronic recognition of pencil markers or printed votes. This class of voting solutions will not be discussed in this paper, where we will focus solely on so-called \emph{direct-recording electronic voting machines} (DRE). These systems do not generate a paper trail\footnote{While \emph{some} DRE systems generate a paper trail, this is only being used in exceptional cases, when the electronic counting can for some reason no longer be relied upon.} and store or transmit the actual vote by means of a digital medium. 

\subsection{Requirements}
\label{sec:requirements}
The following 5 fundamental principles are mandatory for every democratic e-voting solution, according to the council of Europe\cite{eu2004}.
%checked
\begin{itemize}
 \item \textbf{Universal suffrage}: all human beings have the right to vote and to
stand for election, subject to certain conditions such as age and
nationality.
 \item \textbf{Equal suffrage}: each voter has the same number of votes.
 \item \textbf{Free suffrage}: the voter has the right to form and to express his or
her opinion in a free manner, without any coercion or undue influence.
 \item \textbf{Secret suffrage}: the voter has the right to vote secretly as an individual, and the state has the duty to protect that right.
 \item \textbf{Direct suffrage}: the ballots cast by the voters directly determine
the person(s) elected.
\end{itemize}

These can be translated to the following requirements\footnote{Composed from the requirements recommended in \cite{ace}, \cite{hubbers2008} and \cite{elkestem}}:
\begin{itemize}
\item \textbf{Transparency/Integrity}. To make sure that the general public, as well as other key stakeholders in the electoral process, have confidence in the e-voting solution.
\item \textbf{Ballot secrecy/privacy}. To protect the secrecy of the vote at all stages of the
voting process.
\item \textbf{Uniqueness}. To ensure that every vote cast is counted and that each vote is counted only once. 
\item \textbf{Voter eligibility}. To ensure that only persons with the right to vote are able to cast a vote.
\item \textbf{Verifiability/auditing}. Ideally, the voter can check if his vote is counted in the end result. If not, independent auditors should be able to check the integrity of the election result. 
\item \textbf{Accessibility}. To guarantee accessibility to as many voters as
possible, especially with regard to persons with disabilities.
\item \textbf{Vote freedom / coercion resistance}. To maintain the voter’s right to form and to express his or her opinion in a free manner, without any coercion or undue influence.
\item \textbf{Availability}. To ensure that the voting solution is available to the voter during the election.
\end{itemize}

These requirements will be used throughout this paper to judge on voting implementations, and to be able to compare different solutions.

\section{Analysis of risk and benefit}
\label{sec:riskbenefit}
Benefits of e-voting include:
\begin{itemize}
\item{Fast counting}. Counting can be done in mere minutes, compared to hours and sometimes days, depending on the country.
\item{Less labour-intensive}. Although e-voting solutions still requires polling staff, no additional labour is needed for counting.
\item{Cheaper}. A benefit that is often mentioned is that electronic voting will cost less. This however, is often not the case\footnote{The commission researching the reintroduction of electronic voting in the Netherlands concludes that it will cost more\cite{elkestem}}. 
\item{Accessibility}. It is argued that e-voting is more accessible than paper voting for the visual impaired. They can bring headphones and the buttons can be given tactile feedback. On the other hand, for the elderly, e-voting may be more difficult than paper voting.
\end{itemize}
\noindent
Additional benefits of I-voting.
\begin{itemize}

 \item Further improved accessibility. With I-voting, there is no need to go to a polling station. It is more accessible for the physically impaired. It might reduce the amount of people that vote by proxy (which is for example considered a problem in the Netherlands). Lastly, it will be easier for citizens abroad to cast a vote.
\end{itemize}
\noindent
Drawbacks of e-voting:
\begin{itemize}
 \item Scalability of attacks. It is easier to do large scale attacks, because often the same systems and software is being used across a country. In order to influence paper voting counts, an attacker will have to manipulate many different polling stations. This also brings us to the amount of people committing fraud: for e-voting fraud, one or a small group of attackers may be able to change the outcome of an election, while when considering paper elections, in general a larger group is needed.
 \item Less transparency. The e-voting process is a lot less transparent, especially for non-technical people. Advanced knowledge of cryptography is required for people to be able to prove that their vote was taken into account in the election results, and that all the votes were counted correctly. Only a small amount of researchers will comprehend this, while the rest of the population will have to trust a system they cannot understand. 
\end{itemize}
\noindent
Additional drawbacks of I-voting:
\begin{itemize}
 \item Usage on untrusted and unmanaged client systems. I-voting solutions generally assume that the client systems can be trusted. This  assumption does not hold, since many home computers are unsafe. For example, existing botnets could have a significant impact on election results.
 \item Coercion resistance is very difficult if not impossible to achieve. The benefit of voting in a polling station is that the voter goes into the polling booth alone, and does not leave with tangible evidence of the cast vote. If someone attempts to coerce a voter into voting for a certain candidate, there is no way the voter can be expected to prove to them whom he voted for. This does not hold for I-voting. Someone can be forced to vote while others are watching, one can be forced to record a confirmation of their vote, etcetera. 
\end{itemize}

% The International IDEA Policy Paper “Introducing Electronic
% Voting: Essential Considerations” identifies a comprehensive
% range of factors contributing to the building of trust in
% e-voting systems

\section{Traditional paper voting}
To be able to compare paper and e-voting solutions, let us first look at how paper voting satisfies the requirements for a voting solution.

\begin{itemize}% 
\tick \textbf{Transparency}. A paper trail is present, the workings can be explained to every person. A person can be an observer at the voting and counting stage.
\tick \textbf{Ballot secrecy}. Eligibility is checked when entering the polling station, and no identifiable information is present on the ballot. Ballots are folded so they cannot be read and thrown in a box.
\otick \textbf{Uniqueness}. Only one ballot form is provided, but these may be replicated and smuggled in. The amount of double votes which can be cast without getting caught will be very low. Voting twice by visiting different polling stations should be impossible because of voter registration. Voting by proxy might enable an attacker to vote a limited amount of times for others, but requires explicit consent by those who vote by proxy\footnote{In the Netherlands, a signature and a copy of the ID of the person who is voting by proxy is required.}. 
\tick \textbf{Voter eligibility}. To ensure that only persons with the right to vote are able to cast a vote. As is the case with the previous requirement, this is checked by the voting staff. 
\tick \textbf{Verifiability}. With paper voting, procedures are in place to verify the integrity of the voting procedure, and the paper trail allows for reliable verification in case of doubt. 
\ofail \textbf{Accessibility}. Persons with severe disabilities might have difficulty coming to the polling station or filling in the ballot with a pencil.
\otick \textbf{Coercion resistance}. You will not get any proof of what you voted\footnote{There is an attack where a coercer starts with one blank ballot, fills it in, gives a voter the ballot and asks him to put that ballot in the voting bin while taking the empty ballot back to the coercer. The coercer can then fill in this ballot and use it to coerce a next victim.}. A person can however be forced to vote by proxy.
\otick \textbf{Availability}. Paper voting is not dependent on electricity or an internet connection. It is however not very available when there are no polling stations nearby. This might be a problem in rural areas.
\end{itemize}

\section{Case studies electronic voting}
\label{sec:casestudies}

\subsection{I-voting: Estonia}
Estonia is one of the few countries that makes extensive use of I-voting. Introduced in 2005 for local elections, the system has been used ever since and is an increasingly popular way of voting. During the 2015 Parliamentary Elections, 30,5\% of votes cast were I-votes\cite{web:estonia_stats}. It is interesting to note that the voter turnout has slightly increased (61,9\% in 2007, 63,5\% in 2011, 64,2\% in 2015). This supports the often-heard statement that I-voting helps increase voter participation. However, research conducted in Switzerland concludes that``internet voting does not attract new voters or young voters and it has a substitution effect as it replaces postal voting"\cite{doc:swiss-2013}\cite{doc:swiss-2011}. Additional research will be required in order to draw stronger conclusions on how internet voting changes voter participation. 

Until 2013, the Estonian I-voting architecture had not been subjected to a public, independent international review. Eight years after its introduction, Drew Springall and his team were offered a chance to extensively study the Estonian I-voting system\cite{springall2014security}.

\subsubsection{The system}
In order to explain the architecture of the Estonian I-voting system, we will discuss the way a vote is cast, and then follow the vote until the moment it was counted and the election results are published. An analysis of the security of the system is presented in section \ref{sec:estonia-designrisks} while the procedures surrounding the actual voting process will be discussed in section \ref{sec:estonia-proceduralrisks}. 

The architecture of the Estonian I-voting system is as follows. Citizens that want to cast their vote use their web browser to connect to a government website and download a program. This program guides the voter through the voting process, allowing the voter to pick the candidate of his choice. The vote is now encrypted with the \emph{election key}, a public key that is stored in the application and is different for every election. This way, all votes are encrypted and may only be decrypted on a machine that possesses the private election key. Some random data is added to the vote prior to encryption, to make sure no two identical votes result in an identical ciphertext. Omitting this random data would allow an attacker to obtain the contents of the vote by comparing the ciphertext of the vote to all possible encryptions of valid votes. The random data is stored on the client machine for some time, as it is required for vote verification, a mechanism that will be explained shortly. 

The vote now needs to be signed by the voter. This is done with a USB smartcard reader and an Estonian National ID card. This card contains a chip that stores two RSA keys, one for encryption, one for signing, and is widely used for online banking and government services. The RSA keys cannot be extracted and in order to be able to use the card for signatures or encryption, the voter must first enter a pin. The downloaded voting program asks for this pin and then asks the smartcard to sign the (already encrypted) vote. 

The encrypted, signed vote is now sent to the election servers. Upon reception, the election server sends back a unique, unguessable token. The voting program confirms that the vote was submitted, and displays a QR code based on the random data used for the encryption step together with the random token received from the voting server. Using a smartphone app, the voter may scan this QR code. The app then sends the token to the voting server, which responds by sending the encrypted vote. The app now checks if the intended vote plus the encryption randomness indeed encrypts to the same ciphertext as received from the voting server. This verification mechanism is possible for up to 30 minutes, in order to limit the possibility of coercion by others asking for 'proof' that a voter did indeed vote for a specific candidate. 

After reception of the vote by the system, the vote is temporarily stored on the \emph{Vote Forwarding Server}. This server is the only one of the infrastructure that is directly accessible from the internet and verifies voter eligibility before forwarding the vote to the \emph{Vote Storage Server}. 
All votes are collected here until the I-voting phase ends. At that time, when it is no longer possible to cast a vote, all votes that are stored on the vote are once more verified and then the signatures are stripped from the votes. The resulting set of anonymous, encrypted votes are burned onto DVDs. 

The DVDs are then transferred to the Vote Counting Server, an air-gapped machine that contains the election private key. This private key is now used to decrypt and count every vote. The result is the sum of the votes for each candidate.

%Remark: Extensive video footage on Youtube of programming and configuration, partially open source software. 

\subsubsection{Design risks}
\label{sec:estonia-designrisks}
The design of the Estonian I-voting solution has several conceptual shortcomings. These shortcomings are important, as they are very difficult or impossible to mitigate without redesigning important parts of the system. 

\begin{itemize}
\item{Assumes voter's computer is trustworthy\\
This is a dangerous assumption, as many computers are infected with some kind of malware\footnote{Panda Security states that the global infection rate was over 36\% in Q1 2015\cite{pandalabs}, but the accuracy of this estimate is unclear.}. For high-stakes elections, an attacker may be able to attack and infect a large number of computer systems, allowing him to change a vote cast from an infected machine to any candidate of the attacker's choosing. 

Of course, the smartphone-based verification app serves to mitigate the risk of this kind of attacks. However, smartphones are frequently connected to their owners' computers, and an attacker could choose to only modify a vote if an associated smartphone has also been infected. This is not very practical, though, as the attacker will want to modify as many votes as possible in order to maximize his influence in the outcome of the elections. 

A more interesting approach would be to take advantage of the fact that verification is only possible for up to 30 minutes. An attacker could let the user cast his vote, and if after 30 minutes the smartcard is still connected, cast another vote for a candidate specified by the attacker. The user could successfully validate that his vote was cast (as it was indeed properly received by the system), but will not be aware of the fact that his previous vote has been overridden by a new vote cast by the attacker. 
}

\item{Assumes voter can securely download the voting application\\
An attacker controlling any part of the network between the voter and the application download server would be able to intercept and modify any HTTP request/response pair. The download of the application itself is done over HTTPS, which is supposed to provide confidentiality and, more importantly, integrity. However, if an attacker can intercept one of the earlier HTTP requests, he can manipulate the traffic in order to trick the client into downloading the application over an unencrypted socket and provide the user with an infected or patched voting application. 

Another possibility would be for an attacker to obtain a valid certificate for the download server's domain, which should not be possible unless the attacker can convince a certificate authority to cooperate. In this case, the attacker can provide the expected secure HTTPS connection while still tricking the user into downloading a modified application. It is not a simple 
task to convince (or exploit\cite{prins2011diginotar}) a certificate authority in order to obtain a certificate, but assuming state-level or very well funded adversaries this is certainly a scenario to consider. 
}

\item{Assumes the counting server is trustworthy\\
While the counting server is air-gapped and located in a locked rack in the data center, the server is an obvious security risk. If an attacker could in any way tamper with the machine, there is no guarantee that the election results do indeed correspond with the votes cast by the users. There are many ways to tamper with a machine, as malware may be introduced at any point from manufacturing to the moment the votes are being counted. Additionally, some malware may be extremely difficult to detect, such as BIOS-based malware or advanced rootkits. 
}
\end{itemize}

\subsubsection{Procedural risks}
\label{sec:estonia-proceduralrisks}

Part of the security of the Estonian I-voting system relies on procedures, intended to make unauthorized interference impossible. The Internet Voting Committee published extensive procedures, that cover many of the steps and possible events that might occur during the election process. However, there are flaws in the specified procedures, that will be discussed below. 

\begin{figure}
  \centering
  \begin{subfigure}{0.4\textwidth}
    \centering
    \includegraphics[width=170pt]{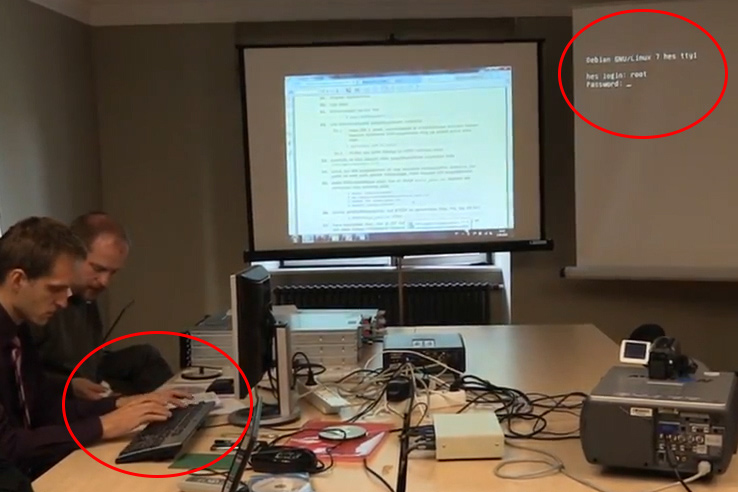}
    \caption{Keyboard filmed when typing in root password.}
    \label{fig:estonia_pic1}
  \end{subfigure}
  \quad
  \begin{subfigure}{0.4\textwidth}
    \centering
    \includegraphics[width=170pt]{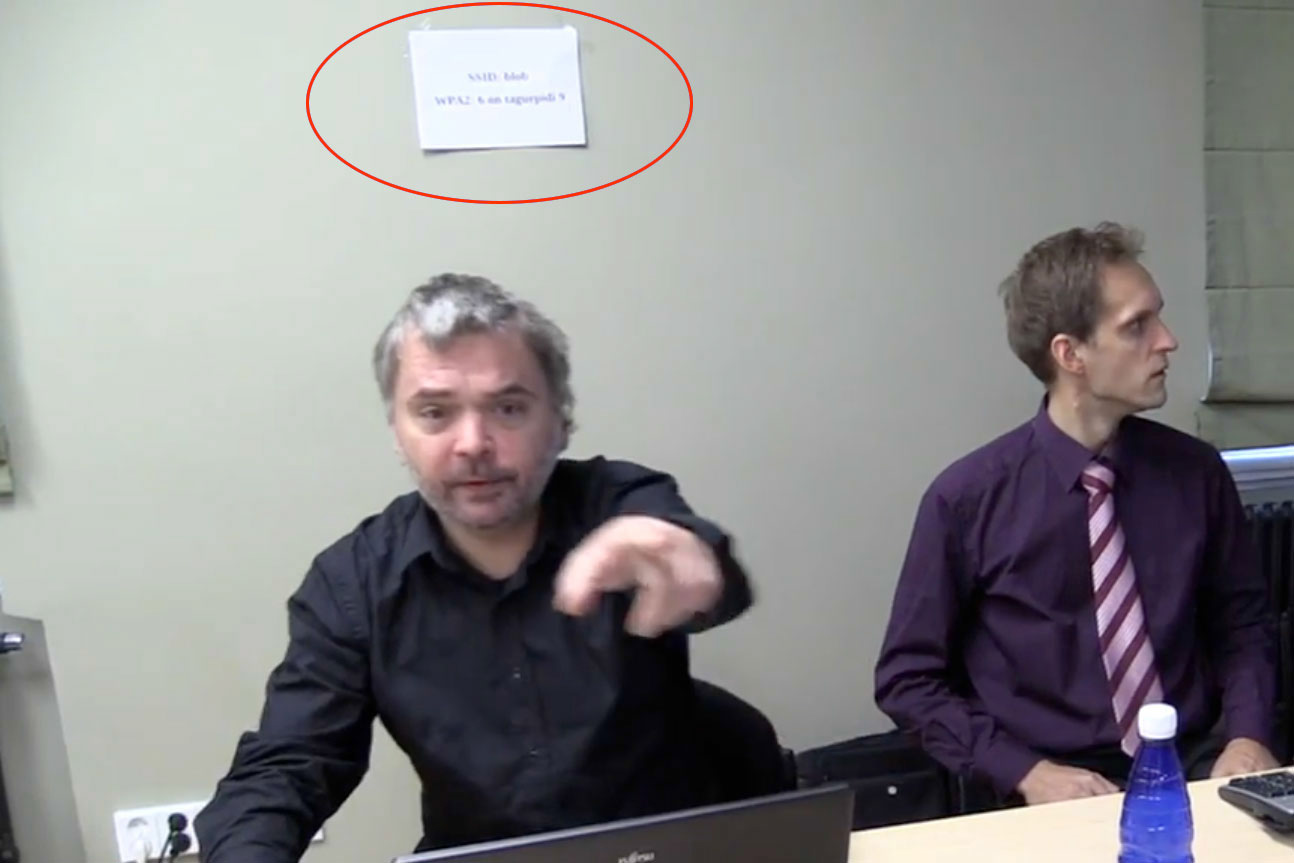}
    \caption{WiFi password readable on a written note on the wall.}
    \label{fig:estonia_pic2}
  \end{subfigure}
  \caption{Source: \cite{springall2014security}}
\end{figure}

\begin{itemize}
\item{Unclear and/or incomplete procedures\\
It was found that some procedures were unclear and/or incomplete. An example of this is that the server racks in the data center are equipped with tamper-evident seals. Although even the added value of these seals is questionable, as they can be defeated with widely available tools\cite{appel2011security}, there was another problem: there was no procedure for how to act if a seal would be found damaged. One might argue that if a situation occurs for which there is no procedure, the integrity of the election can no longer be guaranteed. 

Also, some procedures seemed to be specified or changed during the actual election period. An example is that observers were initially allowed to bring cellphones into the data center, but this changed and was no longer allowed days after the election period started. This suggests that some procedures were not sufficiently thought out. 
}

\item{Leakage of secrets through Youtube videos\\
In an effort to increase the transparency and verifiability of the deployment and configuration of the infrastructure, videos are recorded whenever the system administrators interact with the trusted parts of the system. These videos are then published on Youtube, allowing anyone to verify that the interactions with the system are performed according to protocol and in an honest way. While this is a great idea to mitigate the risk of manipulation by malicious insiders, it also is a security risk, as secret information may unintentionally be recorded and published. Drew Springall et al. found that there are indeed multiple recordings that contain secret information. In figure \ref{fig:estonia_pic1} we see how the root password is being typed in. By carefully analysing the keystrokes, an attacker could either obtain the exact password or at least reduce the search space to a few possibilities per character. Figure \ref{fig:estonia_pic2} shows how a note on the wall shows the password for the local WiFi network. 
}

\item{Use of untrusted computers for sensitive operations\\
Another example is shown in figure \ref{fig:estonia_pic3}, where we see the voting application being compiled. Close inspection shows that the machine used to build the application contains software like \emph{BS Player FREE} and a link to \emph{PokerStars.ee}. This kind of software is not appropriate for a machine that is being used to compile important software, as the presence of malware cannot be excluded.

\begin{figure}
  \begin{center}
  \includegraphics[width=250pt]{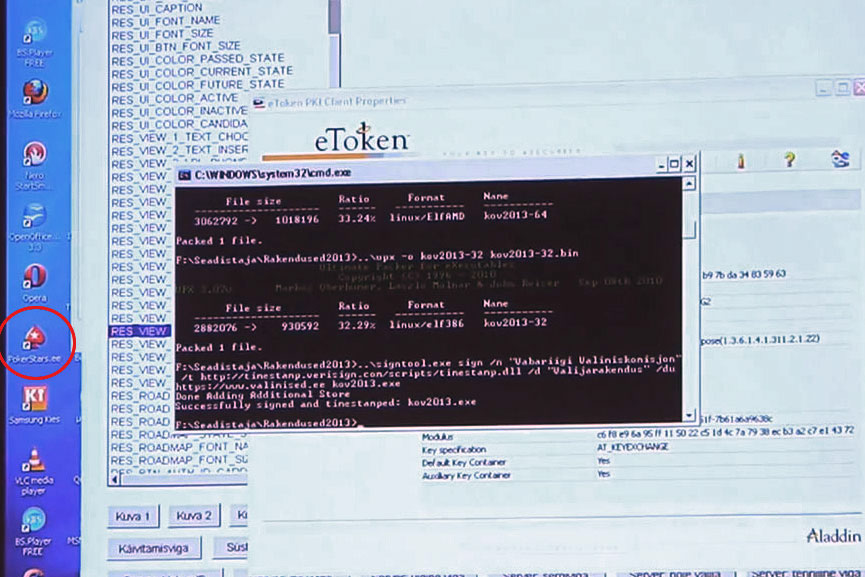}
  \caption{Insecure build machine for voting application. Source: \cite{springall2014security}}
  \label{fig:estonia_pic3}
  \end{center}  
\end{figure} 
}

\item{Occasional deviation from protocol\\
On several occasions, the election staff deviated from protocol. For example, when attempting to burn the encrypted votes to DVD on the Vote Storage Server, there was a technical issue. Instead of resolving the issue in order to be able to burn the votes to DVD according to protocol, a staff member decided to use his personal USB flash drive in order to transfer the votes to the counting server. This is a severe breach of protocol, as the flash drive contained personal documents and was absolutely not guaranteed to be free from malware. More importantly, a flash drive may be crafted by an attacker to exploit some vulnerability on the counting server.
}
\end{itemize} 

\subsubsection{Summary}
When comparing the Estonian I-voting system with the requirements defined in section \ref{sec:requirements}, we come to the following conclusion:

\begin{itemize}
\fail \textbf{Transparency}. Some parts are not open source, some procedures are not on video and one has to trust the counting server. 
\tick \textbf{Ballot secrecy}. Signatures are separated from the votes before counting. Vote contents are only determined afterwards, ensuring anonymity. 
\tick \textbf{Uniqueness}. Voters are allowed to vote multiple times. The backend will check for duplicate votes, only the most recent vote is taken into account. Voting twice will invalidate the first vote. 
\tick \textbf{Voter eligibility}. The national ID card is used for authentication, and voter eligibility is checked properly. 
\fail \textbf{Verifiability}. No independent auditing is done during the election period. Important parts of the system are based on trust. Additionally, deviations from protocol make verification even harder. 
\tick \textbf{Accessibility}. The system is easy to use and does not require physical presence of the voter at a polling station.
\fail \textbf{Coercion resistance}. Although somewhat mitigated by the possibility to vote again, voting from a remote, uncontrolled environment is inherently sensitive to coercion. 
\ofail \textbf{Availability}. A DOS attack has been found, that may disrupt the possibility for electronic voting. However, the impact on the democratic process would be limited as citizens could still go vote in person the day after I-voting ends. 
\end{itemize}

\subsection{E-voting in the Netherlands}

\subsubsection{History and juridicial ground}

There are 5 types of elections in the Netherlands: for the EU, the state, the provinces, the waterboards and the municipalities. %The state has legislative power, and consists of two parts: the house of representatives and the senate. The Netherlands has a multi party system, in which it is very unusual that one party gets a majority of the votes, resulting in coalition governments. 

The Netherlands has had most of its current voting laws since 1922. 
Every person above 18 with Dutch nationality is eligible to vote\footnote{There are some exceptions: a person can be excepted from the right to vote by a judge, prisoners are only allowed to vote by proxy. In January 2009 a law was revoked which denied people with mental disorder the right to vote\cite{art2b}.}. 

Already in 1965, the Dutch electoral act was modified to enable the use of other means than paper voting. The modified law left the choice whether or not to use electronic voting to the municipalities, who are responsible for organizing elections. In 1966, the first voting machine was used\cite{loeber2008}. It was not a success\footnote{It resulted in an abnormal amount of blank votes\cite{hermans2007}}. As a reaction a regulation was created which specified additional requirements for the machines.

In 1989 the law was adjusted to incorporate some requirements, specifically for e-voting machines.
Only machines certified by the ministry of interior may be used. The law\cite{par7} states these minimum requirements\footnote{Other laws also say something about voting machines, like article J 13 - 27 of the ``kiesbesluit", but mostly about procedures: how the list of candidates should be presented, how the election officials shall handle the paperwork, etc. }:
\begin{itemize}
 \item The vote should be secret, even when the voter wishes to cast a blank vote.
 \item The machines should be sturdy and should be able to be used in a simple manner, without malfunctions.
 \item The candidates, the list number and the name of the corresponding party should be presented in a clear way.
 \item The voter should be able to only vote once, but must be able to correct a mistake.
\end{itemize}

%The law does not address (or only very superficially) aspects such as security, privacy and accessibility of the voting process. 
Despite requirements made by TNO\cite{pieters2008} for certifying machines in 1996\cite{voorwaarden}, all kinds of troubles kept haunting electronic voting, including counting errors, inaccessible and unaudited source code and dependence on third parties. Translated from Dutch: 

\begin{quote}
``With the exception of the supplier, no-one knows what the source code is and in which way changes in the program are made. Also, no-one knows how the supplier, Groenendaal bureau for Election results, corrects mistakes.''\cite{hermans2007}
\end{quote}

There is no law which handles auditing during elections or checks on the results of voting with E-voting machines. Since there is no paper trail, no recounting can be done on votes cast on a machine. There are no rules for storage, transportation and the security of the machines. Also the monopoly of Nedap/Groenendaal (bureau for election results) is a worry of the concerning ministers. 
Despite multiple questions, concerns and debates in parliament, although promised otherwise, no big changes were made in the next ten years. 

% In 2004, Nedap, the biggest\footnote{90\% of the used machines were the Nedap machine} of the two E-voting machine makers in the Netherlands, sells 7500 voting computers to Ireland. After research of the company Zerflow questions the basic security of the machine, Ireland decides not to use any of the voting machines. 
% In the Netherlands, a debate is organized in the house of representatives because of this, but no concrete actions are taken.

The Organization for Security and Co-operation in Europe (OSCE) sends a observation mission to the Netherlands in 2006, which criticizes the fact that the machines are closed source, and the lack of transparency caused by the missing paper trail. 
Together with a fraud case in 2006, where a candidate obtained a suspicious amount of preferential votes in the polling station where he was polling worker\footnote{After two appeals Guus te M. is convicted to 180 hours of community service. BC2171 BK9221, ECLI:NL:HR:2010:BK9221, see http://uitspraken.rechtspraak.nl/inziendocument?id=ECLI:NL:HR:2010:BK9221} this sparked the first discussion about voting machines. 

2006 also marks the start of the public debate on the subject of E-voting, fuelled by a group called ``We do not trust voting computers"\footnote{\url{www.wijvertrouwenstemcomputersniet.nl}} (original Dutch name: ``Wij vertrouwen stemcomputers niet"). This group bought a Nedap machine and researched its software and hardware. The conclusions are explained in the next paragraph. The group proved that emitted radiation of the machines destroys vote secrecy(which is a clear requirement by law), and that the physical security of both machine and storage facilities is lacking. 
For the first time, the ministry of interior is confronted with its passive attitude, and the missing extra laws and rules which were promised in the ten years before. The secret service is involved. They conclude that the Dutch voting machines are the least secure compared to machines in other countries. Technical improvements are promised by Nedap and SDU, and the secret service starts helping the ministry by testing the machines for 
radiation emission. This results in decertification of the SDU machines and one of the four types of Nedap machines. 

After the dust settles on the radiation problem, ''we don't trust voting computers" discovers that the software used for counting the votes is not tested against any of the legal requirements\cite{hermans2007}\footnote{Nedap stated they consider the software used for counting not to be a part of the voting machine, and would thus not need to be tested.}. The group starts a court case against the approval of Nedap machines in 2007. As a result all remaining Nedap machines are decertified. A few months later the government announces the withdrawal of the regulation for approval of voting machines. This made it impossible to certify new machines. From this day on the Netherlands again voted with paper ballots.\\

Commissioned by the ministry of general affairs, two advisory committees are formed. One with the goal to find out what went wrong and one to advise on the future of E-voting.

The first committee concludes that the ministry did not have enough technical knowledge of the machines, which caused the companies to control both the market and at the same time be influential in decision making. They also conclude that the legal framework is very weak. Regulation about the voting machine is outdated, and this caused the security of the machines not to be tested properly\cite{hermans2007}.

The second committee recommended a change of responsibility in the electoral process to the minister. It also recommended that there should always be a paper trail, and that a more trustworthy way of using electronic voting is by using a vote printer with a vote scanner, such that counting is improved in time and accuracy, but the system is still similar to the paper ballot system\cite{verkiezingsproces2007}. 

%During the presentation of the report of the second committee, the government announced to withdrawal of the regulation for approval of voting machines. This made it impossible to certify new machines. From this day on the Netherlands again voted with paper ballots.

Despite the controversy, electronic voting never fully disappeared from the agenda\cite{elkestem}. Recently, in February 2015, the ministry of general affairs announced that they want to reintroduce E-voting, now with separate vote printers and scanners.

\subsubsection{Technical implementation}

\paragraph{SDU}
      \begin{figure}
      \centering
      \begin{subfigure}{0.4\textwidth}
      \centering
        \includegraphics[height=100pt]{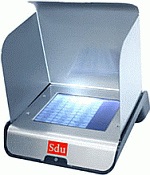}
        \caption{The SDU voting machine}
      \label{fig:sdu}
      \end{subfigure}
      \quad
            \begin{subfigure}{0.4\textwidth}
            \centering
        \includegraphics[height=100pt]{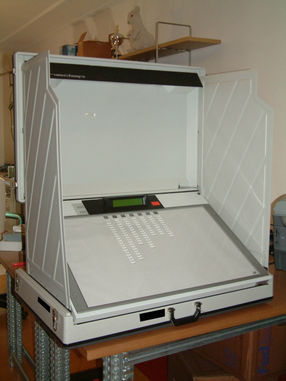}
        \caption{The Nedap voting machine}
\label{fig:nedap}
      \end{subfigure}
\caption{Voting machines in the Netherlands}      
      \end{figure}
      
The SDU voting machines (figure \ref{fig:sdu}) were used in a minority of the municipalities (about 10\% in 2007). Machines are rented, not bought, and delivery, storage, setting up and training is all done by SDU. The SDU machine has a touch screen, which first presents the parties, and after a voter chooses a party, the candidates\footnote{The inability to display all candidates on a screen led to discussion in 1994, and the law had to be changed to enable the use of the machines, for more information see \cite{hermans2007}}.

\paragraph{Nedap}
The majority of machines were made by Nedap (figure \ref{fig:nedap}). Since the failed experiment with American voting machine in 1966, Nedap was involved in the development of voting machines for the Netherlands. The machines have a button for each candidate. It has two simple screens, one for the voter, and one on the election
officials’ console for enabling the machine. The votes are stored in memory in a redundant manner, in arbitrary order\cite{aldini2009}.

\subsubsection{Technical flaws}
We will focus only on the Nedap machines, because they were used more and researched more thoroughly.

\paragraph{Technical shortcomings of the Nedap voting machine}
With the help of the freedom of information act the action group "we do not trust voting computers" obtained information about the voting machines and the process surrounding its approval. They also managed to buy a Nedap voting machine. %They analysed the machine and proved that fraud is possible by writing an program which could transfer votes\footnote{And even play chess}, and they showed that the radiation and sound transmitted by the machine could be used to read out the choice of the voter from a distance (5 meter for nedap, 30 for sdu). 

\begin{figure}
 
   \begin{subfigure}{0.36\textwidth}
   \centering
      \includegraphics[height=90pt]{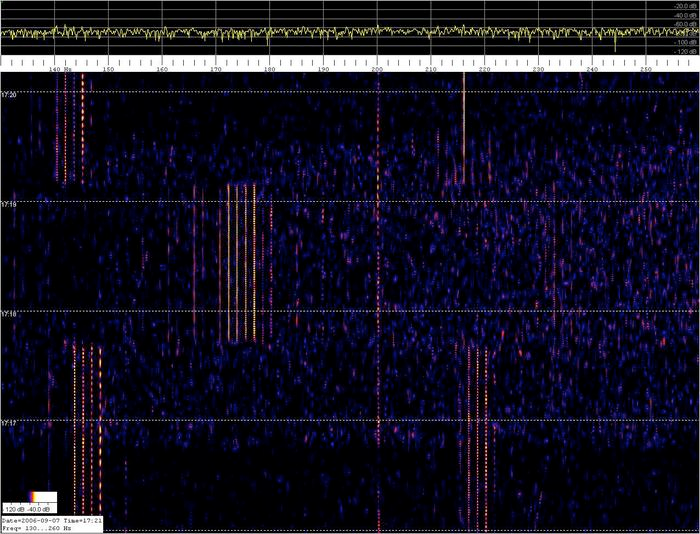}
      \caption{Spectrum waterfall display of the audio signal\cite{gonggrijp2006}.}
      \label{fig:cda}
     \end{subfigure}
        \quad 
     \begin{subfigure}{0.22\textwidth}
     \centering
	\includegraphics[height=90pt]{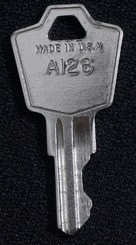}
	\caption{The key of the Nedap machines\cite{gonggrijp2006}.}
	\label{fig:key}       
    \end{subfigure}
    \quad
      \begin{subfigure}{0.36\textwidth}
      \centering
    \includegraphics[height=90pt]{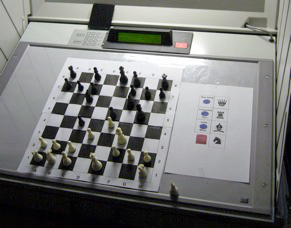}
          \caption{A Nedap machine playing chess\cite{gonggrijp2006}. It was not very good at it.}
\label{fig:chess}
  \end{subfigure}
  \caption{}
\end{figure}

The technical shortcomings of the Nedap/Groenendaal ES3B voting machine brought to light by research of Gonggrijp can be summarized as follows\cite{gonggrijp2006}:
\begin{itemize}
 \item Radio signal emissions: screen signals reveal user vote.
 \item Physical security: same key for every machine, very easy to pick locks, no seals on hardware.
 \item Maintenance mode with password 'geheim' (Dutch for: 'secret').
 \item Ability to install other software (no signature checking)
\end{itemize}
\noindent
\textbf{Radio signal emissions}\\
Gonggrijp et al. found that radio emissions revealed if someone voted CDA (Christian Democrats) or someone else. This was caused by the fact that CDA's name was presented unabbreviated: Christen Democratisch App\'el, which was the only party with a special character (\'e). A different character set had to be loaded, causing the display refresh frequency to drop from 72Hz to 58 Hz (see figure \ref{fig:cda}). This was even audible. Using equipment, the researchers predicted to be able to distinguish all different parties from the emissions, up to a few meters from the machine\footnote{The secret service discovered that this problem was also present in the SDU machines, where the emission revealed a vote up to 30 meters from the machine}.

% \begin{figure}
%    \includegraphics[width=120pt]{./NedapCDA.png}
%    \caption{Spectrum waterfall display of the received audio signal\cite{gonggrijp2006}.}
% \label{fig:cda}
% \end{figure}
\noindent      \\
\textbf{Physical security}\\
Every Nedap machine needs two keys (as required by ``requirements and acceptance for voting machines 1997''\cite{voorwaarden}). It uses a lock system called ``C\&K YL Series 4 Tumbler Camlock''\cite{gonggrijp2006}. This system always comes with the same key(see figure \ref{fig:key}) which meant that every one of the 8000 Nedap voting machines used the same physical key to open and expose the hardware.

% \begin{figure}
%    \includegraphics[width=50pt]{./KeyNedap.png}
%       \caption{The key\cite{gonggrijp2006}.}
% \label{fig:key}
% \end{figure}

\begin{quotation}
``Spare keys can be ordered separately online for roughly an Euro each '...' According to the product datasheet, typical applications for this
lock include ``copy machines and office furniture". The reader unit has, as stipulated by law, a lock with a different key for the slot marked 'programming', which is used to erase the ballot memory modules and to write new candidate lists to the modules. The key is of the same insecure type and the we expect it to also be the same all over the country."\cite{gonggrijp2006}
\end{quotation}
\noindent
Except the use of the same key and the ease with which these can be ordered online, the locks are also extremely sensitive to lockpicking. Every amateur lockpicker would be able to open the locks with ease.\\
\noindent      \\
\textbf{Maintenance mode}\\
The software of the voting machine has a maintenance mode, which should only be accessible by the Nedap helpdesk. You need a password to access it which turned out to be ``geheim", the Dutch word for ``secret''. The maintenance mode allows you to read the binary contents of a ballot module plugged into the programming slot of a reader unit.
\\\noindent \\
\textbf{Installing other software}\\
By sniffing the commands between the software and the reader unit the researchers were able to make a working program on their own, to figure out how the system worked and how and where the candidates and votes were stored. While working on the software, Gonggrijp et al. published on their website that they could manipulate the machine, and that they could just as easily be programmed to play chess as to lie about the election results. In a reaction Nedap dismissed this claim, adding ``[...] And with regard to the claim that our machine can play chess: I’d like to see that demonstrated''. Which, to prove their point, they did (see figure \ref{fig:chess}).

Apart from this they constructed a program which would lie about election results.%, and even be able in some cases to detect if it was tested for fraud.

\subsubsection{Requirements}
When comparing the system with the requirements defined in section \ref{sec:requirements}, we come to the following conclusion:

  \begin{itemize}% 
\fail \textbf{Transparency}. Source code is not open, no paper trail, counting process not transparent. %Nedap/Groenendaal responsible for both the machines and corrections of faults in voting process.
\fail \textbf{Ballot secrecy}. Votes could be derived from electromagnetic radiation and even sound.
\fail \textbf{Uniqueness}. Votes can be manipulated as shown by Gonggrijp et al., which is made easy by a lack of physical security.
\tick \textbf{Voter eligibility}. This is managed by the voting staff, not the voting machines.
\fail \textbf{Verifiability}. No independent auditing of individual verifying possible.
\otick \textbf{Accessibility}. The machines will work comparable with paper voting, if the same color scheme etc. is used. It might be experienced as less accessible by the elderly. Ease of use was also one of the reasons why the fraud case of Guus te M. was possible.
\fail \textbf{Coercion resistance}. Without vote secrecy, you can be coerced to vote for a certain party or candidate, since the coercer can check what is voted. 
\otick \textbf{Availability}. Comparable to paper voting. Only in the case of a power outage does paper voting have an advantage.
\end{itemize}

\subsection{I-voting in the Netherlands}
\label{sec:i-nl}
\subsubsection{History and juridical ground}
As a Dutch citizen, it is possible to vote when living abroad. People who want to do this have to register, and can choose to vote by proxy, or by postal mail. Approximately 25.000 voters register per election to participate\cite{loeber2008}. Voting by postal mail was problematic, so in European parliament elections in 2004, users also had the option to vote by phone of internet.

The system used was called KOA, made by Logica CMG. Out of 15.832 voters, 4.871 voted via the internet\cite{evakoa}\footnote{According to the a document of the ministry of inner affairs. Other sources report 7.195\cite{caarls2010} or 5.335\cite{verkiezingsproces2007} for internet plus telephony votes.}

The experiment was held under special legislation called the ``online voting experiment act''\cite{tekoa}\cite{ekoa}. 

% 
% \begin{wrapfigure}{r}{0.4\textwidth}
% \vspace{-15pt}
% \begin{center}
%      \includegraphics[width=0.35\textwidth]{./2014-Waterschappen-1250.png}%   
%      \end{center}
%      \vspace{-15pt}
% \caption{Water board districts in the Netherlands.}
% \label{fig:waterboards}
% \vspace{-5pt}
% \end{wrapfigure}

The same year a system called RIES (Rijnland Internet Election System) was used for the waterboard elections in two of the 24 waterboards districts.%(area 12 and 19 of figure \ref{fig:waterboards}). 
Voter participation for these election always was fairly low, and internet voting was seen as a means to lower the threshold for people to vote. Against expectations, participation did not increase: out of 1 million eligible voters, only 120.000 voted online.

The third and last experiment was held during the national elections in 2006 for all Dutch citizens living abroad. This time a modified version of RIES was used. Out of 34.305 registered voters 19.815 voted via the internet. 

The discussion on E-voting in 2007 also reached internet voting, and a certifying procedure was announced for the system that November. 

The water boards intended to use RIES for their combined elections in Fall 2008. This was bad timing, as the minister had just stopped the use of electronic voting machines. Parliament demanded an independent evaluation.
SO came that the same year, 3 separate studies\cite{gedrojc2008}\cite{gonggrijp2009}\cite{hubbers2008}) published major flaws in the RIES system, which we will elaborate on in the next paragraphs. This meant the end of internet voting in the Netherlands. 
In 2010 the law\footnote{Experimentenwet Kiezen op Afstand} which facilitated the internet voting experiment was withdrawn.

\subsubsection{Technical implementation}
There are three versions of RIES: RIES-2004 (waterboards), RIES-KOA (EU elections) and RIES-2008 (meant for the waterboards, never used). All studies used the documentation and/or source code of the latest of these 3 versions.\\
\\
Basically, a unique key pair is generated for each voter. This is done before the election. During the election each voter will cast its vote by generating a one-time signature. The signed vote (voted ballot) can be verified once its received by the voting server, or later, when all valid votes are added together. A list with the signed votes and a pseudo-identity is published after the elections so everyone can verify their own vote.%recvote

Assumptions made by RIES include an anonymous channel and a trustworthy client computer.

\subsubsection{Technical flaws}
Three independent parties reviewed RIES. Fox-it, a security company, commissioned by the ministry of infrastructure and environment, based their results on documentation, interviews and the voting website. EIPSI, the Eindhoven Institute for the Protection of Systems and Information, also analysed the system based on the provided documentation. Both studies provide both technical findings and cryptographic findings and both give comments on assumptions made by the builders. Lastly, Gonggrijp et al., based their findings on the source code, which was made available in June 2008. This research was preliminary, done in only a few days, and by no means exhaustive, but does give more technical depth than the other two studies.

Although many more were found, we will mention only five of the most interesting technical findings.

\begin{figure}
      \centering
      \begin{subfigure}{0.4\textwidth}
      \centering
     \includegraphics[width=150pt]{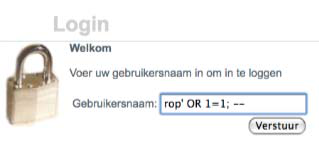}%  
\caption{SQL injection}
\label{fig:sql}
      \end{subfigure}
      \quad
            \begin{subfigure}{0.5\textwidth}
            \centering
	\includegraphics[width=250pt]{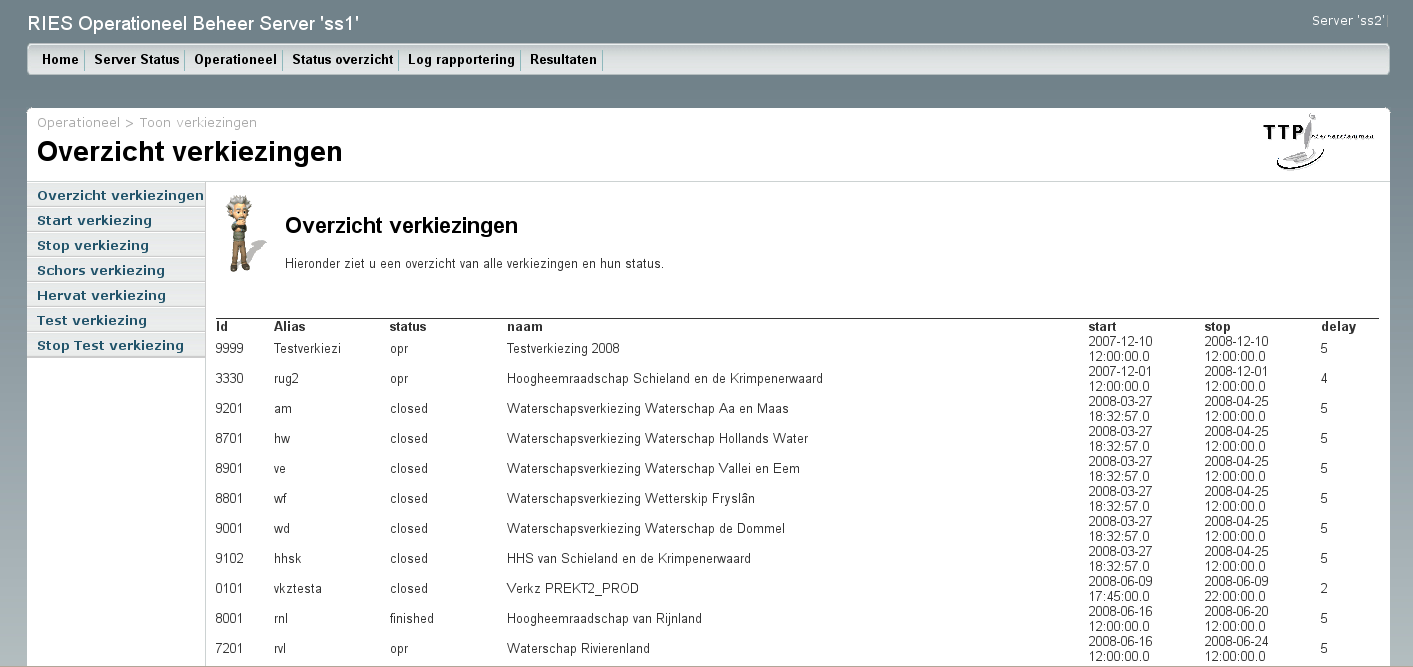}%  
	\caption{Maintenance window}
\label{fig:beheer}
      \end{subfigure}
      \caption{}
      \end{figure}

\begin{itemize} % Gronggrijp
      \item XSS. This vulnerability was actually discovered two years prior, but never mitigated. XSS was possible in the Election ID parameter: \verb|https://stem.surfnet.nl/server/%5C../admin/server?req| \verb|=results&subreq=store&pageid=select&elid=8001<script>alert('XSS')</script>|. %Javascript is vital for the working of the site. It is used for the cryptographic routines that perform the act of voting. 
      \item SQL injection. SQL injection could be used in the username field (figure \ref{fig:sql}) and on the admin panels, where you could for example read the table 'votes'\cite{gedrojc2008}.
      \item The voting station can see cancelled votes. When a vote is cancelled, the chosen candidate and party are sent back to the server as part of the POST parameters.
      \item DoS possible with ';' in URL. A request like \verb|https://stem.surfnet.nl/server/%5C../;images/| would make the server enter an infinite (or at least very long) loop. 
      \item Access to server folders and maintenance windows. See figure \ref{fig:beheer}.
      \end{itemize}   

\paragraph{Cryptographic flaws }

% Authentication is done by using a one-time digital signature scheme\cite{hubbers2008}. DES is used in DESmac and as blockcipher inside MDC-2, 3DES is also used in DESmac, to encrypt sensitive files and for key derivation. DESmac is used for the authentication of voters and ballots. MDC-2 is  hashfunction based on 3DES, used for anonymizing sensite data to enable public verification. RSA is used for wrapping of 3DES keys and certification of public keys and lastly, SHA-1 hashing is used to generate public commitment values ( blz 15 eipso, uitleggen) for public file contents.

~\\~\\In essence, RIES works as follows:
A private key, $Kp$, is generated before the election for every voter. This key generation is not done by the voter, which is usual when using asymmetric cryptography, but centrally, by a company called TTPI, after which the $Kp$ is given to the voter. The key is constructed as follows:\\
$Kp = 2TDES_{Kgenvoterkey} (VnID || ParGp || ElID)$ \\
Where:
\begin{itemize}
   \item $VnID$: a public identity, based on the BSN (the social security number).
   \item $ParGp$: Participant group (to allow for voter groups, in practice only one group is used).
   \item $ElID$: election identifier. %(used for verification) %verkiezingscode
   \item $Kgenvoterkey$: a master key, for one election. This is a 112-bit key.
 \end{itemize}
Note that the same $Kgenvoterkey$, $ParGP$ and $ElID$ is used for all voters in an election, so the diversity of the result only depends on the public identity of the person ($VnID$).

An encrypted vote (called $RnPID$) is constructed as follows:\\
$$RnPID = MDC( DES_{mac_{Kp}} (choice, ElID, birthyear) )$$
%Why the birthyear is added is not explained, and is one of the comments of FOX-it. Although birtyear alone is not enough to identify a person, it is personal data, and there is no reason to use it. The reason given was to prevent abuse of ballot forms, but this is not a very trong argument, since birtyear can be guessed easily.

\textbf{Flaws of this scheme}\\
We again only give a selection of the findings, to illustrate out point. For an exhaustive list, read\cite{gedrojc2008} together with \cite{gonggrijp2009}.

\begin{itemize}
 \item Valid vote codes can be generated. These codes (the $RNPiD$) is generated with a 56 bits DES key. Given that there are in the order of a million voters, and the electionID is known, an attacker needs to generate around $2^{36}$ keys to find a correct one (for the exact computation, see \cite{gedrojc2008}). This would take a normal PC in 2008 about 19 hours. Dedicated hardware would take about 4 seconds.
 \item DES and 3DES. In 2008, DES was already considered a legacy algorithm. The two key variant of 3DES was used for the most important key of the election, the Kgenvoterkey. Because of its short key length, secrecy could not be guaranteed for more than 10 years, and should absolutely not be used in this kind of applications. When an attacker breaks Kgenvoterkey he can:
\begin{itemize}
   \item Determine who was eligible to vote in the election
   \item If (s)he voted
   \item Who (s)he voted for
\end{itemize}
 \item The Printing Service Bureau has the private keys. The PSB handles sensitive data in the clear, in order to print name and address information, together with the private key. %With this information you can vote for someone or check what the person voted. 
 In RIES-2008 the PSB is a foreign, commercial company.\cite{hubbers2008}
 \item Predictable random tokens. The code contains an option of authenticating via mobile phone. When a user wants to log in, he has to echo a random password sent via SMS. The random token takes as a seed the time in milliseconds. It would require an attacker a few thousand guesses to figure out the key. There is no limit on the amount of tries, so this is a feasible attack.
\end{itemize}

 \subsubsection{Requirements}
 When comparing the system with the requirements defined in section \ref{sec:requirements}, we come to the following conclusion:
 \begin{itemize}
\fail \textbf{Transparency}. Source code not open, no independent body.
\fail \textbf{Ballot secrecy}. Not for cancelled votes, not for regular votes within a few years due to poor encryption, not against the printing service bureau.
\tick \textbf{Uniqueness}. Votes can probably be manipulated by XSS, valid voting codes can be calculated, but no evidence was found that these votes can be counted multiple times. 
\fail \textbf{Voter eligibility}. Valid voting codes can be calculated, SQL injection and brute force can be used to bypass authentication.
\fail \textbf{Verifiability}. Much is based on trust, no independent auditing authority.
\fail \textbf{Accessibility}. Works only on some browsers, website does not comply to Dutch accessibility standard for websites.
\fail \textbf{Coercion resistance}. Can never be guaranteed with voting at home, user can make screenshot for his coercer.
\fail \textbf{Availability}. DOS is made easy by infinite loops caused by ';' sign.
 \end{itemize}

\section{E-voting techniques}
\label{sec:improvements}
In this section we discuss two technical proposals that may help mitigate some of the issues that are common for current E-voting and I-voting solutions. These methods have not yet been used in any large-scale E-voting products, but, among other applications, are used by Helios, an on-line I-voting solution that provides strong guarantees about integrity and privacy. After discussing the two concepts, a basic introduction to the Helios system will be presented to show how these concepts can be used in an E-voting solution. 

\subsection{Secret sharing}
Secret sharing is a technique allowing $n$ parties to share a secret, with the property that no single party can recover the secret without cooperation of the other parties. There are also schemes where cooperation of at least $n-a$ parties is required, a so-called \emph{threshold scheme}. A $(k, n)$ threshold scheme divides a secret into $k+n$ shares and would require cooperation of at least $k$ parties in order to recover the secret. 

\begin{figure}[t]
  \begin{center}
  \includegraphics[width=160pt]{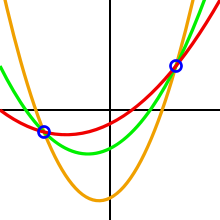}
  \caption{A second-degree polynomial is uniquely defined by three points. Source: Wikipedia\cite{web:wiki-shamirsecretsharing}}
  \label{fig:shamirsecretsharing}
  \end{center}
\end{figure} 

A well known algorithm that implements secret sharing is \emph{Shamir's Secret Sharing}. In this section a basic explanation of such schemes is presented, the interested reader is encouraged to read the wikipedia page on the subject\cite{web:wiki-shamirsecretsharing} or the original paper\cite{shamir1979share}. The algorithm is based on the property that a polynomial of degree $m$ is uniquely defined by $m+1$ points on the curve. The polynomial is considered the secret, and a single point on the polynomial's curve is given to each party. If we have $k$ parties, we need to use a polynomial of degree $k-1$ in order to be able to recover the polynomial. It is easy to extend this scheme to a threshold scheme, by simply generating some more points on the curve and distributing these to the 'extra' parties. This way, one can construct a $(k, n)$ threshold scheme by choosing a polynomial of degree $k - 1$ and generating $k+n$ shares of which each party receives one.\\ 
While the polynomial can only be reconstructed with at least $k$ shares, an attacker that has less than $k$ shares does still have some information about the secret. However, Shamir's Secret Sharing is to be used in a finite field,  which will improve security	 as knowledge of a set of points of size less than $k$ will no longer provide significant knowledge about the secret. 

\subsection{Homomorphic encryption}
Another technique that has very interesting applications in electronic voting schemes is \emph{homomorphic encryption}. A scheme with homomorphic properties allows for computations to be carried out on ciphertext, without the requirement of decryption of the input values and then re-encryption of the result. An encryption scheme is homomorphic if given ciphertexts $\mathcal{E}(a)$ and $\mathcal{E}(b)$ one can obtain $\mathcal{E}(a \perp b)$ for some operation $\perp$. 

In the case of electronic voting, we are interested in the addition operation, allowing us to compute $\mathcal{E}(a+b)$ without decrypting $\mathcal{E}(a)$ or $\mathcal{E}(b)$. This is an interesting property, as it allows to add votes together without the need for decryption of any of the individual votes. This way the counting can be done in a secure and privacy-friendly way, as only the result of the addition has to be decrypted. 

An example of a cryptosystem with the homomorphic property is a modification of ElGamal\cite{elgamal1985public}\cite{wiki:elgamal}. In regular ElGamal, we have public key $h = g^x$ with secret key $x$. The encryption function is $\mathcal{E}(m) = (g^r,m\cdot h^r)$ with random $r$. ElGamal becomes homomorphic under addition when instead of encrypting $m$, one encrypts $g^m$. This variant is called \emph{exponential ElGamal}\cite{cramer1997secure}. Addition of two values $a$ and $b$ under encryption (using random $r_1$ and $r_2$) would be done as follows:

  $$\mathcal{E}(g^{a}) \cdot \mathcal{E}(g^{b})$$
  $$= (g^{r_1}, g^{a}\cdot h^{r_1})(g^{r_2},g^{b} \cdot h^{r_2})$$
  $$= g^{r_1+r_2}, g^{a + b} \cdot h^{r_1+r_2}$$
  $$= \mathcal{E}(g^{a} \cdot g^{b})$$
  $$= \mathcal{E}(g^{a+b})$$

One problem is that when decrypting the end result, one obtains $g^{a+b}$. In order to obtain the actual value of $a+b$, one would have to solve the Discrete Log problem. In this case, this is feasible. Assume we have $n$ votes $(v_0, v_1, .., v_{n-1})$, where each vote is either a $0$ ("no") or a $1$ ("yes"). This implies the following bounds on our sum: 

$$0 \le \sum_{i=0}^{n}(v_i) \le n$$
As the solution space is small, we can simply brute force the different possible sum values in order to find the solution. 

There are also approaches to E-voting that use regular ElGamal encryption, which is homomorphic under multiplication\cite{peng2005multiplicative}. Please note that the above is a highly simplified introduction to homomorphic encryption. In order to design a secure E-voting solution that uses homomorphic encryption, many complications and pitfalls have to be dealt with to mitigate leakage of information and to guarantee integrity\cite{fontaine2007survey}.

\subsection{Helios}
Helios is a system that implements 'voting with cryptographic auditing'. Helios is free, open source software, but can also be used online at the Helios website\cite{web:helios} Because of the nature of the Helios system, trusting the server is not required: the voting process remains fully verifiable even if the system administrators are fully malicious. Helios provides ballot casting assurance, which means that a user can verify that his vote was actually taken into account in the final tally. It also provides universal verifiability, implying that any observer (or user) can check that all votes were counted and tallied correctly. \\
\\
In the paper in which Helios was first presented at Usenix in 2008, the authors state that if a system is fully compromised, a system design can only maintain either integrity, or privacy\cite{adida2008helios}. This first version of Helios was designed to provide strong verifiability guarantees, at the sacrifice of privacy. In a later version, based on a proposal by D. Demirel et al. \cite{demirel2012improving}, the privacy guarantees were strengthened by using homomorphic encryption and multi-party tallying. This was implemented in version 3 of Helios, the version that will be discussed in this section. \\
\\
An important design choice of Helios is that they completely disregard protection against coercion. While other systems go to great length to limit the possibility of coercion, the authors of Helios accept it and even implemented a \emph{coerce me} button, which will send a vote in plain text, accompanied by details required for checking if the vote was indeed cast, to a user supplied email address. The rationale behind this is that by adding such a clear interface for sharing the supposedly ``private'' vote, it is made clear that coercion is to be accepted as a risk that was explicitly not mitigated. \\
\\
A consequence of this design choice is that Helios is not an appropriate choice for high-stakes elections, where interests are such that large-scale coercion might become an effective option for malicious actors. The Helios website\cite{web:helios} states that a classical E-voting (as opposed to I-voting) scenario could be envisioned to limit the risk of coercion. While the focus of this paper is on electronic voting systems suitable for national elections, Helios is being discussed in detail because of the novel features it implements while remaining relatively light-weight. 

\subsubsection{Usage}
In this section we will briefly explain how the Helios voting process works. We assume that an election has already been set up, and that some user Alice wishes to cast a vote. Note that an empty ballot can be viewed and filled in by \emph{anyone}, at any time, without authentication. The authentication (in this context, verification of identity and eligibility for the election) is only done at the moment the ballot is \emph{submitted} to the Helios system. 

\begin{enumerate}
\item Alice selects the election
\item The \emph{Ballot Preparation System} (BPS) leads Alice through the ballot questions and records her choice(s). 
\item After confirmation, the BPS encrypts the answers together with some random data. A hash of ciphertext serves as commitment. 
\item Alice can now choose: audit the ballot, or seal and submit.
\begin{itemize}
	\item If she audits: the ciphertext and key are given to Alice, who can now check if this matches with the commitment and if it decrypts to the vote she wanted to cast. If everything is in order, the BPS continues at step 3 and re-encrypts the choices with new randomness, once more allowing Alice to choose audit or seal. 
	\item If she chooses to seal, all but ciphertext are permanently deleted. The BPS now continues with the next step.
\end{itemize}
\item Alice is asked to authenticate, and if she succeeds, the encrypted vote is registered as Alice's vote. 
\item Helios makes the possibility of coercion explicit by showing a 'coerce me'-button. This button allows Alice to send ciphertext, encryption randomness and plaintext to any email address she specifies. As mentioned before, this is done to explicitly show that coercion resistance is not a property of Helios.
\item Alices encrypted vote now shows up on the bulletin board. Here, all votes that were cast until now are visible. Each vote is either associated to the name of the voter or to some identification number. Anyone who voted can find his encrypted vote registered on the board. Alice can check if her vote indeed shows up and if it was indeed her vote.
\item After the election closes, the election officials\footnote{When using the website, Helios presents itself as a potential election official, but this is not obligatory. All election officials would need to be corrupted in order for the counting integrity to be at risk.} work together in computing the sum of the encrypted votes. This is being done by using secure multi-party computation and homomorphic encryption.
\item The election results are published. Anyone can verify, by using the bulletin board, that his/her vote was taken into account and that the addition of the votes was done properly. 
\end{enumerate}

\subsection{Problems}
While some of the common weaknesses of I-voting solutions have been mitigated, Helios still has some important drawbacks. First, there is the public bulletin board. Anyone can see who (either real identity or pseudonym) has cast a vote. The votes are encrypted, but if the cryptographic algorithms that are used are broken in the future, an attacker would be able to decrypt all votes. Another problem is that while the system could never discreetly modify or remove votes, it can still insert votes for non-existing ``voters'' or for individuals that did not cast a vote. Auditors will have to check the bulletin board for proof that no unexpected identities are listed, while some people who did not vote will have to check that indeed no vote was registered under their name. Lastly, there is the vulnerability to coercion that the authors of Helios consider inherent to I-voting. 

\subsection{Summary}
When comparing the system with the requirements defined in section \ref{sec:requirements}, we come to the following conclusion:

\begin{itemize}
\tick \textbf{Transparency}. The design of the Helios system is fully open source and allows for end-to-end verifiability.
\otick \textbf{Ballot secrecy}. All encrypted votes are visible on the bulletin board. While the votes remain encrypted even when counting, the encrypted votes may be broken if the encryption algorithm is to be broken in the future. 
\tick \textbf{Uniqueness}. The bulletin board allows for no more than one vote associated with an identity. 
\tick \textbf{Voter eligibility}. The bulletin board allows only votes that are associated with an eligible (pseudonymous) identity.
\tick \textbf{Verifiability}. Helios offers end-to-end verifiability and one can check if the counting was done properly. 
\tick \textbf{Accessibility}. Helios is accessed over the internet. 
\fail \textbf{Coercion resistance}. Helios explicitly does not protect against coercion and even implements a \emph{coerce me} button. 
\fail \textbf{Availability}. Helios is accessible over the internet, which makes it vulnerable to denial of service attacks. 
\end{itemize}

\section{Conclusion}
\label{sec:conclusion}
Considering the current shift from paper forms to their digital, on-line counterparts, replacing paper voting by electronic voting systems seems to be a logical step. However, the actual advantages of electronic voting over paper voting are more limited than some suggest. Electronic voting will lead to faster vote counting, allowing for election results to be published earlier. Also, it will reduce the amount of wasted paper and, at least theoretically, reduce the amount of invalid votes (due to improper selection or unreadable ballots). It will generally not, however, result in cheaper elections, as estimates show that electronic elections tend to be more expensive than regular paper elections. \\
\\
When considering internet voting, the equation changes quite drastically. While internet voting has the advantages that it greatly increases the accessibility of elections and potentially \emph{is} cheaper than paper elections, important additional risks are associated. The cost argument is largely invalidated by the fact that, in general, internet voting does not fully replace regular paper voting; instead, voters have the choice to either vote from home or in person. Omitting the possibility of voting in person would put the availability at risk, as denial of service attacks are a distinct possibility. For high stakes elections, the assumption that the voter's computer system is not infected by malware (or at least, that no malicious program will interfere with the voting process) is, in our eyes, unacceptable. Well-funded adversaries may be able to infect a percentage of the population and modify or block a part of the votes that are cast. Lastly, coercion resistance is not a realistic property for 
internet voting, as social pressure and the potential presence of others when voting may lead to different voting behaviour. The possibility of casting a new vote at a later time does not fully mitigate this.\\
\\
For both E-voting or I-voting it holds that it is very hard to guarantee the integrity of the election procedure. Computers are extremely complex and many different attack vectors exist. Although the techniques described in section \ref{sec:improvements} might help mitigate a part of this problem, these solutions are extremely complex, and it remains impossible for a regular citizen to understand what guarantees can be given on the integrity of the election results. As we are dealing with a key part of the democratic process, the authors of this paper judge that the benefits of electronic voting solutions do not outweigh the risks. 

\bibliographystyle{plain}
\bibliography{citations.bib}

\end{document}